\documentclass[multphys,vecphys]{svmult}

\usepackage{makeidx}     
\usepackage{graphicx}    
\usepackage{multicol}    

\makeindex             

\begin{document}

\title*{On the Stability of Thermonuclear Burning Fronts in Type Ia Supernovae}
\author{F.~K.~R{\"o}pke\inst{1}\and
W.~Hillebrandt\inst{2}}
\institute{Max-Planck-Institut f{\"u}r Astrophysik,
  Karl-Schwarzschild-Str.~1,\\ D-85741 Garching, Germany\\
$^1$\texttt{fritz@mpa-garching.mpg.de}
$^2$\texttt{wfh@mpa-garching.mpg.de}}
%
%
\maketitle

\begin{abstract}The propagation of cellularly stabilized thermonuclear
flames is investigated by means of numerical simulations. In Type Ia
supernova explosions the corresponding burning regime establishes at scales
below the Gibson length. The cellular flame stabilization---which is a
result of an interplay between the Landau-Darrieus instability and a
nonlinear stabilization mechanism---is studied for the case of
propagation into quiescent fuel as well as interaction with vortical
fuel flows. Our simulations indicate that in thermonuclear supernova
explosions stable cellular flames develop around the Gibson scale and that a
deflagration-to-detonation transition is unlikely to be triggered from
flame evolution effects here.
\end{abstract}

\section{Introduction}
\label{sec:1}

The standard model of Type Ia supernovae (SNe Ia) describes
these astrophysical events as thermonuclear explosions of white dwarf
(WD) stars \cite{hoyle1960a}. In our study, we refer to the specific
scenario (for a review on SN Ia explosion models see
\cite{hillebrandt2000a}), where the
WD consists of carbon and 
oxygen and the thermonuclear reaction propagates in form of a flame
that
starts out in the so-called \emph{deflagration mode}. That is, the flame
is mediated by microphysical transport processes and its burning
velocity is subsonic. One key ingredient in such a SN Ia model is the
determination of the effective propagation velocity of the
deflagration flame. The so-called laminar burning speed $s_l$, i.e.~the
propagation velocity of a \emph{planar} deflagration flame, is much
too low (a few percent of the sound speed in the unburnt material) to
explain powerful SN Ia explosions. The solution to this problem is
provided by the concept of turbulent combustion. Instabilities on
large scales
result in the formation of a \emph{turbulent cascade}, where
large-scale eddies decay into smaller
ones thereby transporting kinetic energy from large to small
scales. Interaction of the flame with those eddies wrinkles the flame
front and enlarges its surface. This is equivalent to an increase in
the net fuel consumption rate and hence causes an acceleration of the flame.

Recent SN Ia explosion models on scales of the WD could show that
in this way enough energy can be released to gravitationally unbind the star
\cite{reinecke2002d,gamezo2003a}. However, these models cannot resolve
all relevant length scales down to the flame width and therefore have to
rely on assumptions on the physics on unresolved small scales. The
goal of our small-scale simulations of the propagation of the
thermonuclear flame is to test those assumptions and eventually to reveal
new physics that additionally needs to be included in the SN Ia
models. The most significant assumptions of the large scale models by
Reinecke et al.~ (e.g.~\cite{reinecke2002d}) are that the energy input
into the turbulent cascade
originate solely from large scales and the flame be stable on
unresolved scales.

A fundamental feature of the turbulent cascade is that in a certain
intermediate scale range, the turbulent velocity fluctuations decrease
monotonically with smaller length scales of the eddies. This is
captured by the corresponding scaling law. From that effect it is
obvious,
that there must exist a cutoff scale, below which flame propagation is
not affected by the turbulent cascade anymore. Below this \emph{Gibson
  scale}, the velocity fluctuations are so small, that the flame burns
faster through the eddies than they can deform it. Thus the flame here
propagates through ``frozen turbulence''. Our simulations aim on the
study of effects around and below the Gibson scale. Except for the
very late stages of the SN Ia explosion, the Gibson scale is
well-separated from the width of the flame and the flame may safely be
regarded as a discontinuity between burnt and unburnt material.

\section{Theoretical background}

What are the effects that dominate flame propagation below the Gibson
scale? Here, two counteracting effects become important. The first is
the \emph{Landau-Darrieus} (LD) \emph{instability}
\cite{darrieus1938a,landau1944a}. It is based on a
hydrodynamical effect and would completely prevent flame stability on
the scales under consideration. This would have
drastic impact on SN Ia models. 
However, there exists a second effect of purely geometrical origin
\cite{zeldovich1966a}, which balances the LD
instability. Once the perturbations have grown to a critical size,
former recesses of the flame front will develop into cusps, which
possess propagation velocities slightly exceeding the burning speed of
the rest of the front.
This effect gives rise to a
stable cellular shape of the
flame. Therefore, the regime of flame propagation that will be studied
in the present contribution is termed \emph{cellular burning regime}.

In connection with
SN Ia explosions, this burning regime was studied by
\cite{niemeyer1995a,blinnikov1996a} and \cite{roepke2003a}
demonstrated its existence by means of a full hydrodynamical
simulation. The questions that arise here are: Can the cellular flame
stabilization break down under certain conditions (e.g.~low fuel
densities or interaction with turbulent velocity fields)? If so, could
the flame itself generate turbulence on small scales (in contradiction
to the assumption of large-scale SN Ia models) and thereby actively
accelerate? This conjecture was put forward under the name
\emph{active turbulent combustion} by \cite{niemeyer1997b} on
the basis of \cite{kerstein1996a}. The questions are closely
connected to the search for a mechanism providing a
deflagration-to-detonation transition (DDT) of the flame propagation
mode in the context of the \emph{delayed detonation model} of SNe
Ia. A detonation wave is mediated by shocks and propagates with
supersonic velocities.

\section{Results of numerical simulations}

In order to find answers to the above questions, we performed
numerical simulations of the flame propagation in two dimensions. For
a description and tests of the numerical implementation we refer the reader to
\cite{roepke2003a,roepke_phd}. Our simulations addressed the flame propagation
into quiescent fuel as well as flame interaction with a vortical flow
field. The simulations were performed on an equidistant cartesian
grid. The flame was
initialized in the computational domain and perturbed in
a sinusoidal way from its planar shape. To follow the
long-term flame evolution, we performed the simulations in a frame
of reference comoving with the flame. This was implemented by imposing
an inflow boundary condition ahead of the flame front and an outflow
boundary on the opposite side of the domain. Transverse to the
direction of flame propagation we applied periodic boundary
conditions.

\subsection{Flame propagation into quiescent fuel}

\begin{figure}[t]
\centering
\includegraphics[width=0.69 \textwidth]{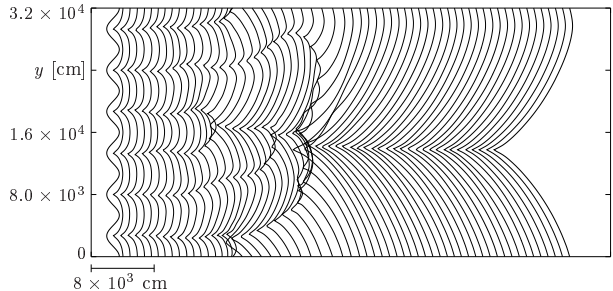}
\caption{Flame propagation into quiescent fuel of a density of $5
  \times 10^7 \,\mathrm{g}\,\mathrm{cm}^{-3}$. Each contour marks a
  time step of $3.3 \times 10^{-3} \, \mathrm{s}$. The contours are
  shifted for better visibility and the spacing between them does not
  reflect the flame velocity.}
\label{evo_q_fig}
\end{figure}

The first part of our investigations addressed the flame propagation into
quiescent fuel. Here two stages of flame evolution are revealed. In
the beginning, the small sinusoidal perturbation imposed on the flame
front increases due to the LD instability. This part of flame
evolution was studied in \cite{roepke2003a}, were it was shown that
the growth rates observed in our simulations agree well with the
expectations from Landau's linear stability analysis
\cite{landau1944a}. 
In the second, nonlinear stage the flame stabilizes in a cellular pattern.
The propagation of a flame with an initial
perturbation of 6 periods fitting into the domain is plotted in
Fig.~\ref{evo_q_fig}.
The tendency of the small cells to merge in the nonlinear regime
finally stabilizing in a single domain-filling cusp-like structure is
apparent here. This result is consistent with semi-analytical studies
of flame evolution \cite{gutman1990a}.

\subsection{Flame propagation into vortical fuel}

In SN Ia explosions, 
turbulent velocity fluctuations around the Gibson scale can  be
expected and also relics from pre-ignition
convection may contribute to turbulent flows.
In order to explore the effects of flame interaction with such velocity
fluctuations, we modified the inflow boundary condition to a so-called
\emph{oscillating} inflow condition, which generates vortices by
imposing the following velocities at the boundary
\begin{eqnarray*}
v_x &=& s_l \left\{ -1 + v' \sin 2 k \pi y \cos 2 k \pi (x - t
  s_l)\right\}\\
v_y &=& s_l\, v' \cos 2 k \pi y \sin 2 k \pi (x - t
  s_l).
\end{eqnarray*}
Here, the parameter $v'$ characterizes the strength of the imprinted
velocity fluctuations and $k$ denotes the wavenumber of the
oscillation.
This produces what is termed ``square vortices'' by
\cite{helenbrook1999a}.

\begin{figure}[t]
\centering
\includegraphics[width= 0.7 \textwidth]{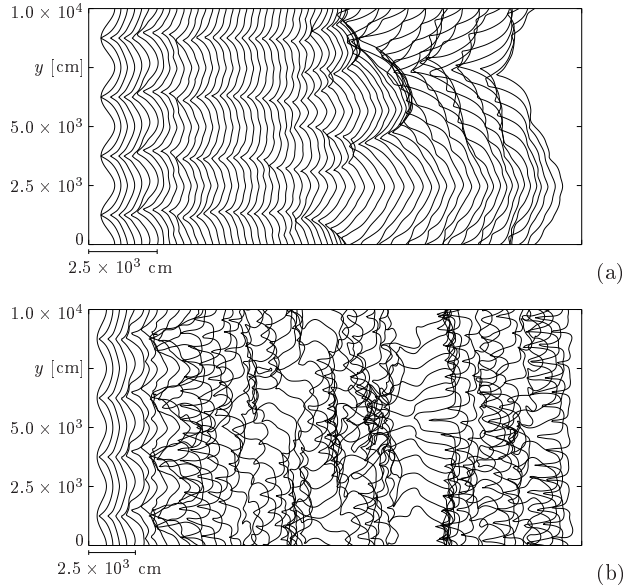}
\caption{Flame propagation into vortical fuel of a density of $5
  \times 10^7 \,\mathrm{g}\,\mathrm{cm}^{-3}$. \textbf{(a)} $v'/s_l =
  0.7$; \textbf{(b)}
  $v'/s_l = 2.5$. Each contour represents a
  time step of $8.0 \times 10^{-4} \, \mathrm{s}$ (a), and $2.4
  \times 10^{-3} \, \mathrm{s}$ (b). Again, the contours are
  artificially shifted for better visibility.}
\label{evo_v_fig}
\end{figure}

The results of two of such simulations for different strengths of the
square vortices are plotted in Fig.~\ref{evo_v_fig}. 
Although in case of
weak vortices in the incoming flow there is some interaction with the
flame, the flame still shows the tendency to align in a
large-wavelength cellular structure (cf.~Fig.~\ref{evo_v_fig}a). This
is similar to the case of flame propagation into quiescent fuel. On
the other hand, if the vortices are strong enough, they can completely
break up the cellular stabilization and prevent the flame from
evolving into a single domain-filling structure
(cf.~Fig.~\ref{evo_v_fig}b). However, we do not observe a drastic
increase in flame surface and thus flame propagation speed. The flame
structure rather adapts to the vortices imprinted on the fuel flow.

\section{Conclusions}

The presented simulations of flame propagation into quiescent fuel are
in good agreement with theoretical
expectations. The linear stage of flame evolution is consistent with
Landau's dispersion relation \cite{roepke2003a, landau1944a} and in
the nonlinear regime the flame stabilizes in a cellular pattern. Thus,
our hydrodynamical model of flame evolution shows that the cellular
burning regime exists for SN Ia explosions.

Interaction of the flame with vortical flow fields may lead to a
break-down of cellular stabilization if the velocity fluctuations are
sufficiently large. In this case, however, the flame shows the
tendency to adapt to the imprinted flow field. Thus, we observe a
moderate increase in the effective flame propagation velocity but no
drastic effects. No indication of active turbulent combustion could
be found. Therefore, it seems unlikely that effects around the Gibson
scale account for a DDT at late stages of the SN Ia explosion as has
been anticipated by \cite{niemeyer1997b}. More detailed discussions
of flame propagation into quiescent fuel at different fuel densities
and interaction with vortical flow fields of varying strengths are
currently in preparation \cite{roepke2003b,roepke2003c}.



\end{document}